\documentstyle[12pt]{article}
\def\double{\baselineskip 24pt \lineskip 10pt}
\def\l{\label}
\newcommand{\be}{\begin{equation}}
\newcommand{\ee}{\end{equation}}
\newcommand{\bea}{\begin{eqnarray}}
\newcommand{\eea}{\end{eqnarray}}
\newcommand{\ar}{\alpha_r(k)}
\newcommand{\br}{\beta_r(k)}
\newcommand{\k}{\vec{k}}
\newcommand{\kk}{\tilde{k}}
\newcommand{\mr}{\mu_{(r)}}
\newcommand{\ms}{\mu_{(r-1)}}
\newcommand{\er}{\eta_r}
\newcommand{\vae}{\varepsilon}
\newcommand{\md}{\frac{q_r}{|q_r|}}

\newcommand{\amd}{\frac{q_{r-1}}{|q_{r-1}|}}
\newcommand{\amod}{\frac{|q_{r-1}|}{q_{r-1}}}
\newcommand{\1}{{\sf P}^{m_{r}}_{\nu_{r}}(X_r)}
\newcommand{\2}{{\sf Q}^{m_{r}}_{\nu_{r}}(X_r)}
\newcommand{\3}{{\sf P}^{l_{r}}_{n_{r}}(Y_r)}
\newcommand{\4}{{\sf Q}^{l_{r}}_{n_{r}}(Y_r)}
\newcommand{\5}{{\sf P}^{(m_{r})+1}_{\nu_{r}}(X_r)}
\newcommand{\6}{{\sf Q}^{(m_{r})+1}_{\nu_{r}}(X_r)}
\newcommand{\7}{{\sf P}^{(l_{r})+1}_{n_{r}}(Y_r)}
\newcommand{\8}{{\sf Q}^{(l_{r})+1}_{n_{r}}(Y_r)}
\newcommand{\A}{{\sl P}^{m_{r}}_{\nu_{r}}(\overline{X}_r)}
\newcommand{\B}{{\sl Q}^{m_{r}}_{\nu_{r}}(\overline{X}_r)}

\newcommand{\D}{{\sl Q}^{l_{r}}_{n_{r}}(\overline{Y}_r)}
\newcommand{\E}{{\sl P}^{(m_{r})+1}_{\nu_{r}}(\overline{X}_r)}
\newcommand{\F}{{\sl Q}^{(m_{r})+1}_{\nu_{r}}(\overline{X}_r)}

\newcommand{\I}{{\sl Q}^{(l_{r})+1}_{n_{r}}(\overline{Y}_r)}
\begin{document}
\begin{titlepage}
\pagestyle{empty}
\begin{flushright}
{\footnotesize Brown-HET -1047\\
May 1996}
\end {flushright}
\begin{center}
\Large
{\bf GRAVITON PRODUCTION IN ELLIPTICAL AND HYPERBOLIC
UNIVERSES}\\
\vspace{.2in}
\normalsize
\large{M. R. de Garcia Maia$^{(a),}$\footnote{E-mail:
mrgm@dfte-lab.ufrn.br} and
J. A. S. Lima$^{(a),(b),}$\footnote{E-mail: 
limajas@het.brown.edu}}\\
\normalsize
\vspace{.2in}
{\em $^{(a)}$Departamento de F\'{\i}sica\\Universidade Federal do Rio 
Grande do Norte\\ 59072-970 Natal RN Brazil}\\
\vspace{.1in}
{\em $^{(b)}$Department of Physics, Brown University\\Providence RI 02912
U.S.A.}
\end{center}
\vspace{.1in}
\begin{abstract}
\noindent
\baselineskip=24pt
The problem of cosmological graviton creation for homogeneous and isotropic
universes with elliptical ($\vae =+1$) and hyperbolical ($\vae =-1$) geometries
is addressed. The gravitational wave equation is established  for a 
self-gravitating fluid satisfying the barotropic 
equation of state $p=(\gamma -1)\rho$, which is the source of the Einstein's 
equations plus a cosmological $\Lambda$-term. The time dependent part of this 
equation is exactly solved in terms of hypergeometric functions for any value
of $\gamma$ and spatial curvature $\vae$. An
expression representing an adiabatic vacuum state is then obtained in terms of 
associated Legendre functions whenever $\gamma\neq \frac{2}{3}\; 
\frac{(2n+1)}{(2n-1)}$, where n is an integer. This includes most
cases of physical interest such as $\gamma =0,\;4/3\;,1$.
The mechanism of graviton creation is reviewed and the Bogoliubov coefficients
related to transitions between arbitrary cosmic eras are also explicitly 
evaluated.

\vspace*{.3cm}
PACS number(s): 04.30.Db, 09.62 +v, 98.80.Hw 
\end{abstract}
\end{titlepage}
\double
\baselineskip 0.6cm
\section{Introduction}
\l{s1}
\def\theequation{\thesection.\arabic{equation}}
\setcounter{equation}{0}

After the pioneer works of Grishchuk \cite{gri}, the formation of a 
gravitational wave (GW) background in the course of the cosmological expansion
has been extensively studied by several authors (see, for example, 
\cite{gw3t} and references therein).  These `Grishchuk gravitons' \cite{carr}
are created because the definition of particle states at early and late times
is not the same and hence an initial vacuum state seen later becomes a 
multi-particle state \cite{gw3t},\cite{gw1}--\cite{hu93}. 
The direct detection of these
waves would provide us with a new view of the very early phases of the universe
since any graviton should have decoupled from matter and from other fields as
early as the Planck time ($10^{-43}s$) \cite{lukashSta}--\cite{matzner88}. 
In principle, even without
detecting them directly, we may obtain useful information and constrain
important parameters of any cosmological models by analyzing their indirect 
effects. In fact, besides probing the current view of the very early universe,
the presence of a GW background can, for example, cause deviations
from the Hubble flow for the velocities of galaxies and clusters of galaxies
\cite{carr}, affect the timing of orbital motions \cite{carr} and pulsars 
\cite{stine},
influence the results of primordial nucleosynthesis \cite{gw3t,gw1}, and,
through the Sachs-Wolfe effect \cite{sWol}, may also contribute to the 
anisotropies of the Cosmic Microwave Background (CMB) \cite{allenK94,gri94}.

In spite of the importance of the subject, the large majority of the
studies on graviton production found in the literature deal only with the case 
of  spatially flat ($\vae =0$) 
Friedmann-Robertson-Walker (FRW) universes. In fact,
to the best of our knowledge, only in two articles the time dependent part of
the GW equation is solved explicitly for non-spatially flat FRW cosmologies,
namely:
in the original work of Lifshitz \cite{lif}, where the solution is given for
radiation- and matter-dominated universes with hyperbolic geometries 
($\vae =-1$), and in the recent paper by Allen, Caldwell and Koranda 
\cite{allenCk95}
where the authors have considered  an elliptical ($\vae =+1$) FRW universe and 
a sequence of a de Sitter (inflationary) phase, followed by a 
radiation-dominated era and a matter-dominated (dust) phase. In ref. 
\cite{fordP77} the GW equation for $\Lambda=0$ and
any $\vae$ was established, but not solved
for $\vae \neq 0$, whereas in \cite{griClo} the problem of long GW's in a 
homogeneous closed universe (type $\rm G_3 \;IX$ in the Bianchi classification)
was addressed (see also \cite{abS86}).

In the present paper we generalize previous results by studying the creation of
gravitons in both elliptical ($\vae =+1$) and hyperbolic ($\vae =-1$) FRW 
cosmologies. For these values of $\vae$,
we derive the general solution of the GW equation  in a model where the 
density $\rho$ and pressure $p$ of
the cosmic fluid obey the equation of state $p=(\gamma -1)\, \rho$, $\gamma\leq
2$. We find the solution representing an adiabatic vacuum state \cite{birFul}
for a large class of values of $\gamma$ ($\gamma\neq \frac{2}{3}\; 
\frac{(2n+1)}{(2n-1)}$, $n=0,\;\pm 1,\;\pm 2, \ldots$). By considering an 
arbitrary
sequence of cosmic eras with $\gamma$ in this class, we evaluate the Bogoliubov
coefficients which are necessary to obtain the graviton spectrum 
\cite{gw3t,gw1} and to give an accurate determination of their contribution
to the anisotropy of the CMB \cite{allenK94}. This exact result implies that it
is possible to perform a direct comparison among the stochastic background of
GW's produced in different FRW 
geometries ($\vae =0, \;\pm 1$), and potentially it may also be used 
to impose constraints on several cosmological parameters (the case $\vae =0$ was
treated in detail in \cite{gw3t,gw1}).

The paper is organized as follows: In Sec. 2 we establish the wave equation
for any value of $\vae$. In Sec. 3 the cosmological model considered is
described and the general solution for the scale factor in terms of the 
conformal time, $\eta$, is given. In Sec. 4 we derive the general solution for
the GW equation in terms  of hypergeometric functions, whereas in Sec. 5 we
obtain the solutions representing an adiabatic vacuum as a combination of
associated Legendre functions. The mechanism of graviton creation is 
reviewed in Sec. 6 and the Bogoliubov coefficients are evaluated for the model
under consideration. A brief summary of our results is presented in a conclusion
section.

Unless otherwise stated, the units used are such that $\hbar=c=k_b=1$. Greek
indices run from $0$ to $3$ (spacetime indices) and latin indices from $1$ to
$3$ (spatial indices).
\section{The wave equation}
\l{s2}
\def\theequation{\thesection.\arabic{equation}}
\setcounter{equation}{0}

The study of primordial GW's is done through the use of a perturbation
formalism developed by Lifshitz \cite{lif}. This consists in making an expansion
in the deviations from an unperturbed, non-radiative configuration, and 
linearizing the Einstein equations about that configuration \cite{thor}. The
spacetime metric is
\be
\l{eq1}
ds^2=g_{\mu\nu}^{(T)}\,dx^{\mu}\,dx^{\nu}\;,
\ee
with
\be
\l{eq2}
g_{\mu\nu}^{(T)}=g_{\mu\nu}+h_{\mu\nu}\;,
\ee
\be
\l{eq3}
h_{\mu\nu}\ll g_{\mu\nu}\;,
\ee
where  $h_{\mu\nu}$ are small perturbations to the background metric 
$g_{\mu\nu}$ which can be decomposed into scalar, vector, and 
transverse-traceless tensor modes. The first two modes represent density and
rotational perturbations, respectively. The latter transforms as a 3-tensor
under spatial coordinate  transformations, and corresponds to sourceless, weak
GW's. We, therefore, consider only perturbations of the Einstein equations such
that the perturbed value of the energy-momentum tensor, $T_{\mu\nu}$, is set 
equal to zero (see Eq. (\ref{eq20}) below).

For a homogeneous and isotropic universe, and in the co-moving coordinate 
system, the background line element takes the FRW form
\be
\l{eq4}
ds^2=dt^2-a^2(t)\,dl^2=a^2(\eta)\,(d\eta^2-dl^2)\;,
\ee
where $t$ and $\eta$ are, respectively, the cosmic and conformal times, related
by
\be
\label{eq5}
dt=a\,d\eta\; .
\ee
The 3-space metric is 
\bea
\l{eq6}
dl^2&=&\hat{g}_{ij}\,dx^i\,dx^j\\ 
\l{eq7}
 &=&\frac{1}{1-\vae r^2}\, dr^2 + r^2\, (d\theta^2+\sin^2\theta\, d\varphi^2)\\
\l{eq8}
 &=&d\chi^2+{\cal F}^2(\chi)\, (d\theta^2+\sin^2\theta\, d\varphi^2)\;,
\eea
\bea
\l{eq9}
{\cal F}(\chi)=\frac{\sin(\sqrt{\vae}\,\chi)}{\sqrt{\vae}}\,   &=&
\sin\chi\hspace{3.1mm},\hspace{3mm}\vae=+1 \nonumber\\ 
 &=&\chi\hspace{9.1mm},\hspace{3mm}\vae=0 \nonumber\\ 
 &=&\sinh\chi\hspace{2.2mm},\hspace{3mm}\vae=-1\;.
\eea
We shall use the form of $dl^2$ in terms of $\chi\;,\theta\;,\varphi$ 
with $0\leq\theta\leq\pi$, $0\leq\varphi <2\pi$, 
$0\leq\chi\leq\pi$ for $\vae =+1$, and $0\leq\chi<\infty$ for $\vae =0,\:-1$.
Note that
\be
\l{eq10}
g_{ij}=-a^2(t)\hat{g}_{ij}\;.
\ee

The Einstein equations read
\be
\l{eq11}
R_{\mu\nu}-\frac{1}{2}\,R\,g_{\mu\nu}+\Lambda\,g_{\mu\nu}=-8\pi G\,T_{\mu\nu}\;,
\ee
with $R_{\mu\nu}$, $R$, and $\Lambda$ being the Ricci tensor, the Ricci scalar,
and the cosmological constant, respectively.
For a perfect fluid the energy-momentum tensor is
\be
\l{eq12}
T_{\mu\nu}=(\rho+p)\,U_{\mu}\,U_{\nu}-p\,g_{\mu\nu}\;,
\ee
and $U^{\mu}$ is the four-velocity of the fluid. The field equations take the
form
\be
\l{eq13}
\frac{8\pi G}{3}\,\rho\,+\,
\frac{\Lambda}{3}=\frac{\dot{a}^2}{a^2}\,+\,\frac{\vae}{a^2}\;,
\ee
\be
\l{eq14}
8\pi\, G\, p - \Lambda=
-2\,\frac{\ddot{a}}{a}\,-\,\frac{\dot{a}^2}{a^2}\,-\,\frac{\vae}{a^2}\;,
\ee
where a dot means derivative with respect to the cosmic time $t$. By assuming 
that the equation of state is
\be\l{eq15}
p=(\gamma-1)\,\rho\hspace{3mm},\hspace{3mm}\gamma\leq 2\;,
\ee
we get from (\ref{eq13}) and (\ref{eq14})
\be\l{eq16}
\rho=\rho_0\,\left (\frac{a_0}{a}\right )^{3\gamma}\;,
\ee
\be\l{eq17}
a\ddot{a}+\frac{3\gamma-2}{2}\,\left (\dot{a}^2+\vae\right)\,-
\frac{1}{2}\,\gamma\,\Lambda\,a^2=0\,.
\ee
As one may check, the above equation has the first integral
\be\l{eq18}
\dot{a}^2=\left (a_0^2H_0^2+\vae +\frac{1}{3}\Lambda\,a_0^2\right )
\,\left (\frac{a_0}{a}\right )^{3\gamma-2}\,-\vae
+\frac{1}{3}\Lambda\,a^2\;,
\ee
where the Hubble parameter is defined by
\be\l{eq19}
H(t)=\frac{\dot{a}}{a}=\frac{a^{'}}{a^2}\;,
\ee
the primes represent derivatives with respect to $\eta$, and 
$\rho_0 \equiv \rho(t_0)$, etc.

To obtain the GW equation we perturb the Einstein equations to first order and
set
\be\l{eq20}
\delta\rho=\delta p=\delta U^{\mu}=0\;,
\ee
\be\l{eq21}
U^{\mu}h_{\mu\nu}=0\;.
\ee
We also impose the gauge conditions
\be\l{eq22}
\nabla_{\nu}h^{\mu\nu}=0\;,
\ee
where $\nabla_{\nu}$ indicates the covariant derivative. Eqs. (\ref{eq21}) and
(\ref{eq22}) lead to
\be\l{eq23}
{h^{\mu}}_{\mu}=0\;.
\ee
(When $a(t)=$ constant, (\ref{eq22}) constitutes only 3 independent conditions
and (\ref{eq23}) must be imposed separately \cite{fordP77}.) In the coordinate
system of (\ref{eq4}) $U^{\mu}=\delta^{\mu}_{0}$, and Eq. (\ref{eq21}) gives
\be\l{eq24}
h_{0\mu}=0\;,
\ee
and from (\ref{eq22})
\be\l{eq25}
\hat{\nabla}_j\,h^{ij}=0\;.
\ee
($\hat{\nabla}_j$ represents the covariant derivative with respect to the 
spatial metric $\hat{g}_{ij}$.) We are then left with only two independent
components of $h^{\mu\nu}$, corresponding to the two polarizations of a GW.

In the gauge used we obtain \cite{fordP77}
\be\l{eq26}
\frac{d^2 {h_i}^j}{dt^2}+3\, H\, \frac{d{h_i}^j}{dt}+ \frac{2\vae}{a^2}\,
{h_i}^{j}=\frac{1}{a^2}\,\hat{\nabla}^2\,{h_i}^{j}\;,
\ee
where
\be\l{eq27}
\hat{\nabla}^2\,{h_i}^{j}\equiv \hat{g}^{lm}\,\hat{\nabla}_l\,\hat{\nabla}_m\,
{h_i}^{j}\,=\,\frac{1}{\sqrt{\hat{g}}}\,\frac{\partial}{\partial x^l}\,\left
(\sqrt{\hat{g}}\,\hat{g}^{lm}\,\frac{\partial {h_i}^j}{\partial x^m}\right )\;,
\ee
and $\hat{g}$ is the determinant of $\hat{g}_{ij}$. In terms of the conformal
time, Eq. (\ref{eq26}) can be recast as
\be\l{eq28}
({h_i}^j)^{''}\,+\,2\,\frac{a^{'}}{a}\,({h_i}^j)^{'}\,+\,2\vae\,{h_i}^j=
\hat{\nabla}^2\,{h_i}^j\;.
\ee

If we write
\be\l{eq29}
\vec{x}=(\chi,\,\theta,\,\varphi)\;,
\ee
then it is possible to find mode solutions of (\ref{eq26}) (or (\ref{eq28})),
labeled by $\kk$, in the form
\be\l{30}
{h_i}^j(\kk,\,t,\,\vec{x})=\Psi(k,\,t)\,{G_i}^j(\kk,\,\vec{x})\;,
\ee
where the functions ${G_i}^j(\kk,\,\vec{x})$ and $\Psi(k,\,t)$ satisfy
\bea
\l{eq31}
\hat{\nabla}^2{G_i}^j(\kk,\,\vec{x})&=&-(k^2-3\vae)\,{G_i}^j(\kk,\,\vec{x})\;,\\
\l{eq32}
\hat{\nabla}_j{G_i}^j(\kk,\,\vec{x})&=&0\;,\\
\l{eq33}
{G_i}^i(\kk,\,\vec{x})&=&0\;,
\eea
\be\l{eq34}
\ddot{\Psi}(k,t)+3\,\frac{\dot{a}}{a}\,\dot{\Psi}(k,t)+(k^2-\vae)\,\frac{\Psi
(k,t)}{a^2}=0\;,
\ee
or still, in the conformal time
\be\l{eq35}
\Psi^{''}(k,\eta)+2\,\frac{a^{'}}{a}\,\Psi^{'}(k,\eta)+(k^2-\vae)\,\Psi(k,\eta)
=0\;.
\ee

In the above equations, $k$ is the co-moving wave number, related to the 
physical wavelength $\lambda$ and frequency $\omega$ by
\be\l{eq36}
k=\frac{2\pi a}{\lambda}=\omega\,a\;,
\ee
and \cite{lif}
\be\l{37}
0<k<\infty\hspace{9mm},\hspace{3mm}\vae=0,\,-1\;,
\ee
\be\l{eq38}
k=3,\,4,\,\ldots\hspace{3mm},\hspace{3mm}\vae=+1\;.
\ee
For $\vae=0$, the symbol $\kk$ stands for
\be\l{eq39}
\kk\equiv (K;\k)\;,
\ee
where $K$ represents one of the two possible polarization states of the GW's, i.
e., $K=K_a$ or $K=K_b$; $\k$ is the co-moving wave vector,
\bea
\l{eq40}
\k&=&(k_1,k_2,k_3)\hspace{3mm},\hspace{3mm}-\infty <k_j <\infty\;,\\
\l{eq41}
k&=&|\k|=\left (\sum_{i=1}^3\,k_i^2\right )^{1/2}\;.
\eea

For $\vae=\pm 1$,
\bea
\l{eq42}
\kk &\equiv &(K;\,k,J,M)\;,\\
\l{eq43}
M&=&-J,\,-J+1,\,\ldots,\,J\;,\\
\l{eq44}
J&=&2,\,3,\,\ldots,\,k-1\hspace{3mm},\hspace{3mm}\vae=+1 \nonumber\\ 
 &=&2,\,3,\,\ldots\hspace{15mm},\hspace{3mm}\vae=-1\;.
\eea
(Note that `our' $k^2$ coincides with the $\beta^2$ of \cite{abS86}, but is
$3\vae$ plus the $k^2$ of ref. \cite{fordP77}, and that `our' $k$ is $1$ plus 
the $L$ of ref. \cite{allenCk95}.)

It is convenient to express the temporal dependence of the wave in terms of the
function $\mu(k,t)$ defined by
\be\l{eq45}
\Psi(k,t)=\frac{\mu(k,t)}{a(t)}\;.
\ee
Eqs. (\ref{eq34}) and (\ref{eq35}) lead, respectively, to
\be\l{eq46}
\ddot{\mu}(k,t)+\,\frac{\dot{a}}{a}\,\dot{\mu}(k,t)+\left [\frac{(k^2-\vae)}
{a^2}-\left (\frac{\dot{a}}{a}\right )^2-\frac{\ddot{a}}{a}\right ]\,\mu(k,t)=0
\;,
\ee
\be\l{eq47}
\mu^{''}(k,\eta)+\left (k^2-\vae-\frac{a^{''}}{a}\right )\,\mu(k,\eta)=0\;.
\ee
We shall take Eq. (\ref{eq47}) as the basic wave equation to be solved.

The general solution representing the tensor perturbations can therefore be
written as
\be\l{eq48}
{h_{\mu}}^{\nu}(\eta,\vec{x})=\sum_{\kk}\left [A(\kk)\,\Psi(k,\eta)
{G_{\mu}}^{\nu}(\kk,\vec{x})\, +\, {\rm H.C.}\right]\;,
\ee
where H.C. represents the Hermitian conjugate of the previous term, and $\sum$
refers to summation over discrete values and integration over continuous ones, 
i. e. \cite{fordP77,parkF},
\bea\l{eq49}
\sum_{\kk}&=&\sum_{K}\:\int d^3k=\sum_{K}\:\int_0^{\infty}\,k^2\,dk\hspace{3mm},
\hspace{3mm}\vae=0\nonumber\\
 &=&\sum_K\:\sum_{k,J,M}\hspace{37mm},\hspace{3mm}\vae=+1\nonumber\\
 &=&\sum_K\:\int_0^{\infty}\,dk\:\sum_{J,M}\hspace{24mm},\hspace{3mm}\vae=-1\;,
\\
\l{eq50}
\sum_K&=&\sum_{K=K_a}^{K_b}\;,\\
\l{eq51}
\sum_{k,J,M}&=&\sum_{k=3}^{\infty}\:\sum_{J=2}^{k-1}\:\sum_{M=-J}^J\hspace{9mm},
\hspace{3mm}\vae=+1\;,\\
\l{eq52}
\sum_{J,M}&=&\sum_{J=2}^{\infty}\:\sum_{M=-J}^J\hspace{17mm},\hspace{3mm}\vae=-1
\;.
\eea

In the coordinates used, the spatial functions are specified by Eqs. 
(\ref{eq31})--(\ref{eq33}) and by
\be\l{eq53}
{G_{\mu}}^0=0\;.
\ee
We shall impose the normalization condition
\be\l{eq54}
\int\,{G_{\mu}}^{\nu}(\kk,\vec{x})\,{G_{\mu}^{*}}^{\nu}(\kk^{'},\vec{x})\,
\sqrt{\hat{g}}\,d^3x=16\, \pi\, G\, \delta(\kk,\kk^{'})\,
\ee
where the asterisk indicates the complex conjugate and $\delta(\kk,\kk^{'})$ is
the delta function with respect to $\sum_{\kk}$ \cite{parkF}, i. e.,
\be\l{eq55}
\sum_{\kk}g(\kk^{'})\,\delta(\kk,\kk^{'})=g(\kk)\;.
\ee
It proves convenient to choose the phases of the ${G_{\mu}}^{\nu}(\kk,\vec{x})$
in such a way that 
\be\l{eq56}
{G_{\mu}^{*}}^{\nu}(\kk,\vec{x})={G_{\mu}}^{\nu}(-\kk,\vec{x})\;,
\ee
where
\bea
\l{eq57}
-\kk&=& -(K;\,\vec{k})=(K;\,-\vec{k})\hspace{23.1mm},\hspace{3mm}\vae=0\;,\\
\l{eq58}
-\kk&=& -(K;\,k,J,M)=(K;\,k,J,-M)\hspace{3mm},\hspace{3mm}\vae=\pm 1\;.
\eea
(See \cite{parkF} for details.)

In the classical theory $A(\kk)$ and $A^{\dagger}(\kk)$ in Eq. (\ref{eq48}) are
just complex constants (Fourier-Bessel coefficients),
whereas in the quantized theory they become annihilation
and creation operators, respectively. As usual, these operators are required to
satisfy the commutation relations
\bea
\l{eq61}
[A(\kk), A^{\dagger}(\kk^{'})]&=&\delta(\kk,\kk^{'})\;,\\
\l{eq62}
[A(\kk), A(\kk^{'})]&=&0\;.
\eea
We further impose the following 
normalization condition on the Wronskian of the time-dependent part of the 
unified solution
\be\l{eq59}
\Psi(k,\eta)\,\Psi^{*'}(k,\eta)\,-\,\Psi^{*}(k,\eta)\,\Psi^{'}(k,\eta)\, =
\,\frac{i}{a^2}
\ee
or, equivalently,
\be\l{eq60}
\mu(k,\eta)\,\mu^{*'}(k,\eta)\,-\,\mu^{*}(k,\eta)\,\mu^{'}(k,\eta)\, =\,i\;.
\ee
%
\section{The solution of the field equations}
\l{s3}
\def\theequation{\thesection.\arabic{equation}}
\setcounter{equation}{0}
\subsection{General equations}

The problem of determining the spectrum of primordial GW's in a consistent way
may be summarized in four basic steps:
\begin{enumerate}
\item First of all one finds the $a(\eta)$ for the cosmological
model under consideration by solving (\ref{eq17}) or, equivalently, 
(\ref{eq18}).
This is required in order to obtain the explicit
form of the wave equation (\ref{eq47}). Particular solutions of
the field equations have usually been considered  in the literature. 
In this work a {\em general} treatment will be presented.
\item The solution of (\ref{eq47}) should then be found. As remarked earlier,
some approximate methods have been used in the literature (see, for example, 
\cite{sahni90}).
\item By studying the behavior of the solutions in the high-frequency limit,
the {\em particular} solution of (\ref{eq47}) representing
an adiabatic vacuum state \cite{birFul} is identified.
\item Finally, the Bogoliubov coefficients are evaluated using, for instance, 
the general procedure described in \cite{gw3t}.
\end{enumerate}

In this section we deal with the first step by writing the solutions in a
convenient way for our purposes. Following the procedure developed in ref.
\cite{assadL}, we initially recast (\ref{eq17}) using the conformal time 
\cite{jb87}
\be\l{eq63}
\frac{a^{''}}{a}\,+\,\frac{(3\gamma-4)}{2}\,\left (\frac{a^{'}}{a}\right )^2\,
+\,\frac{(3\gamma-2)}{2}\,\vae\,-\,\frac{1}{2}\,\gamma\,\Lambda\,a^2\,=\,0\;.
\ee
For $\gamma=2/3$ we make the substitution
\be\l{eq64}
b=\ln a
\ee
to get
\be\l{eq65}
b^{''}-\frac{1}{3}\,\Lambda\,e^{2\,b}=0\;,
\ee
which can be easily solved for $\Lambda=0$. The scale factor is then
\be\l{eq66}
a(\eta)=a_0\,e^{a_0\,H_0\,(\eta-\eta_0)}\hspace{3mm},\hspace{3mm}\gamma=\frac{2}
{3}\hspace{3mm},\hspace{3mm}\Lambda=0\;.
\ee
For $\gamma\neq 2/3$, by replacing
\bea
\l{eq67}
b&=&a^{1/q}\;,\\
\l{eq68}
q&\equiv &\frac{2}{3\gamma-2}\;,
\eea
Eq. (\ref{eq63}) takes the following form
\be\l{eq69}
b^{''}+\frac{\vae}{q^2}\,b-\frac{\gamma\,\Lambda}{2q}\,b^{2\,q+1}=0\;,
\ee
which formally is the nonlinear equation satisfied by a forced harmonic
oscillator.

From now on, we restrict our analysis to models with vanishing cosmological
constant. In this case, (\ref{eq69}) reduces to the equation of a
simple harmonic oscillator, whose solution is
\be\l{eq70}
b=c_1\sin \left(\sqrt{\frac{\vae}{q^2}}\,\eta+c_2\right )\hspace{3mm},
\hspace{3mm}\gamma\neq \frac{2}{3}\hspace{3mm},\hspace{3mm}\Lambda=0\;,
\ee
and from (\ref{eq67}), we write the unified expression for the scale factor as
\be
\l{eq71}
a(\eta)=a_0\left(\frac{\sin\Theta}{\sin\Theta_0}\right)^q\hspace{3mm},
\hspace{3mm}\gamma\neq \frac{2}{3}\hspace{3mm},\hspace{3mm}\Lambda=0\;,
\ee
where,
\bea
\l{eq72}
\Theta&\equiv &\frac{\sqrt{\vae}}{|q|}\,(\eta-\eta_0)\,+\,\Theta_0\;,\\
\l{eq73}
\Theta_0&\equiv &\Theta(\eta_0)={\rm arctan}\left (\frac{q}{|q|}\,\frac{a_0H_0}
{\sqrt{\vae}}\right)\;,
\eea
and by definition
\bea
\l{eq74}
a_0&\equiv &a(\eta_0)\;,\\
\l{eq75}
H_0&\equiv &H(\eta_0)\;.
\eea
Note also that the unified expression for the Hubble parameter is given by
\be\l{eq76}
H(\eta)=\frac{q}{|q|}\,\frac{\sqrt{\vae}\,\cot\Theta}{a}\;.
\ee
>From (\ref{eq13}) with $\Lambda=0$ and 
\be\l{eq77}
\rho_c=\frac{3H^2}{8\,\pi\,G}\;,
\ee
the density parameter is then
\be\l{eq78}
\Omega\equiv\frac{\rho}{\rho_c}=1+\frac{\vae}{a^2H^2}=
\frac{1}{\left(\cos\Theta\right)^2}\;.
\ee

In the limit $\vae\rightarrow 0$, (\ref{eq71}) reduces to the general
form of the solution for the spatially flat case (Eq. (28) of ref. \cite{gw1}).
Since the graviton spectrum for $\vae=0$ was studied in detail in 
\cite{gw3t,gw1}, in what follows we shall treat only the cases $\vae=\pm 1$.
\subsection{Elliptical models ($\vae =+1$)}

In these models, $\Theta$ and $\Theta_0$ are real and, for $q>0$ ($\gamma >2/3$,
non-inflationary scenarios), $\Theta_0 = {\rm arctan}(a_0H_0)$, whereas for
$q<0$ ($\gamma <2/3$, inflationary scenarios), $\Theta_0= \pi -{\rm arctan}(a_0
H_0)$. Therefore, for any $q$, $0<\Theta_0 <\pi$ and $\sin\Theta_0 >0$. For
$a(\eta)$ to be positive we must then require $0<\Theta <\pi$. 

For $q>0$ 
($\ddot{a}<0$) there is an initial singularity at $\Theta=0$ ($\eta_s =\eta_0-
q\Theta_0$). An expansion phase occurs for $0<\Theta <\pi /2$ until $a(\eta)$ 
reaches the maximum value 
$a_{max}=a_0/(\sin\Theta_0)^q$ at $\Theta=\pi /2$ ($\eta_{max}
=\eta_0+q\left(\frac{\pi}{2}-\Theta_0\right )$). This is followed by a 
contraction stage for $(\pi/2)<\Theta <\pi$, and finally by a `big-crunch' at 
$\Theta=\pi$ ($\bar{\eta}_s=\eta_0+q(\pi-\Theta_0)$).

For $q<0$ ($\ddot{a}>0$), $a$ is arbitrarily large as $\Theta\rightarrow 0$
($\eta\rightarrow\eta_{\infty_{1}}=\eta_0-|q|\Theta_0$), decreases for $0<
\Theta <\pi/2$, reaches the minimum $a_{min}=a_0(\sin\Theta_0)^{|q|}$
at $\Theta=\pi/2$ ($\eta_{min}=\eta_0+|q|\left(\frac{\pi}{2}-\Theta_0\right )$),
grows for $(\pi/2)\,<\Theta <\pi$, and becomes arbitrarily large as $\Theta
\rightarrow\pi$ ($\eta\rightarrow\eta_{\infty_{2}}=\eta_0+|q|(\pi-\Theta_0)$).

Note that $\cos\Theta$ is positive for $\left(0<\Theta <
\frac{\pi}{2}\;,\;q>0\right)$, $\left(\frac{\pi}{2}<\Theta <\pi\;,\;q<0\right)$,
and negative otherwise.
\subsection{Hyperbolical models ($\vae =-1$)}

For $\vae=-1$ we define
\bea
\l{eq79}
\Phi&\equiv &\frac{\Theta}{i}=\frac{1}{|q|}\,\left[(\eta-\eta_0)+q\,\Phi_0
\right]\;,\\
\l{eq80}
\Phi_0&\equiv &\frac{|q|}{q}\,\frac{\Theta_0}{i}={\rm arcth}\,(a_0H_0)\;,
\eea
where $|a_0H_0|>1$. Hence,
\bea
\l{eq81}
a(\eta)&=&a_0\,\left[\frac{\sinh\Phi}{\sinh\left(\frac{q}{|q|}\,\Phi_0\right)}
\right]^q=a_0\,\left[\frac{|q|}{q}\,\frac{\sinh\Phi}{\sinh\Phi_0}\right]^q\;,\\
\l{eq82}
H&=&\frac{q}{|q|}\,\frac{\coth\Phi}{a}\;,\\
\l{eq83}
\Omega&=&\frac{1}{\left(\cosh\Phi\right)^2}\;.
\eea
For an expanding universe we must have
\be\l{eq84}
a_0H_0>1\;.
\ee
Consequently, $\Phi_0>0$, $\sinh\Phi_0>0$. The condition $a>0$ then requires
$\Phi>0$ ($\eta>\eta_0-q\,\Phi_0$) for $q>0$, and $\Phi<0$ ($\eta<\eta_0+|q|\,
\Phi_0$) for $q<0$.

For $q>0$ there is a singularity at $\Phi=0$ ($\eta_s=\eta_0-q\,\Phi_0$) and
$a\rightarrow\infty$ as $\Phi\rightarrow\infty$ ($\eta\rightarrow\infty$).
For $q<0$, $a\rightarrow 0$ as $\Phi\rightarrow -\infty$ ($\eta\rightarrow -
\infty$) and $a\rightarrow\infty$ as $\Phi\rightarrow 0$ ($\eta\rightarrow
\eta_0-q\,\Phi_0$).
\section{General solution for the wave equation and the effective potential}
\l{s4}
\def\theequation{\thesection.\arabic{equation}}
\setcounter{equation}{0}
For the cosmological model described in Sec. \ref{s3}, the wave equation 
(\ref{eq47}) reads 
\be\l{eq85}
\mu^{''}(k,\eta)+(k^2-\vae-a_0^2H_0^2)\,\mu(k,\eta)=0\hspace{15.5mm},
\hspace{3mm}\gamma=\frac{2}{3}\;,
\ee
\be\l{eq86}
\mu^{''}(k,\eta)+\left[k^2+\frac{(1-q)}{q}\,\frac{\vae}{(\sin\Theta)^2}\right]
\,\mu(k,\eta)=0\hspace{3mm},\hspace{3mm}\gamma\neq\frac{2}{3}\;.
\ee
The solution of (\ref{eq85}) is trivial and we shall concentrate only in the 
case $\gamma\neq\frac{2}{3}$. 

Note that, as it happens in the spatially flat geometry, the effective 
potential \cite{gw3t,gw1,gw2} 
\be\l{eq87}
V_{eff}=\left |\frac{(1-q)\,\vae}{q\,(\sin\Theta)^2}\right |
\ee
approaches zero as $\gamma\rightarrow 4/3$. If we define the Hubble co-moving
wave number by
\be\l{eq88}
k_H\equiv\frac{2\,\pi\,a}{\lambda_H}=\frac{2\,\pi\,a}{H^{-1}}\;,
\ee
and 
\be\l{eq89}
S\equiv\frac{V_{eff}}{k_H^2}\;,
\ee
we see that the ratio $S$ is time dependent, i. e.,
\be\l{eq90}
S(\Theta(\eta))=\frac{1}{4\,\pi^2}\,\left |\frac{1-q}{q}\right |\,\frac{1}
{(\cos\Theta)^2}\;.
\ee
Therefore, the amplification condition \cite{gri,gw3t,gw1,gw2}
\be\l{eq91}
k^2\ll V_{eff}
\ee
does {\em not} necessarily represent modes outside the Hubble radius, $\lambda
\gg\lambda_H$ ($k^2\ll k_H^2$). In fact, this will happen only for those time
intervals where $S(\Theta(\eta))<1$. In elliptical geometries, however, we may 
have $S(\Theta(\eta))>1$ as long as
\be\l{eq92}
-\frac{1}{2\,\pi}\,\sqrt{\left |\frac{1-q}{q}\right|}\,<\cos\Theta\,<\,
\frac{1}{2\,\pi}\,\sqrt{\left |\frac{1-q}{q}\right |}\;,
\ee
that is, close to $a_{max}$ for $q>0$ and close to $a_{min}$ for $q<0$. As
$|\cos\Theta|\leq 1$, (\ref{eq92}) is restricted to hold for $q$ in one of
the following intervals: $(1+4\,\pi^2)^{-1}\leq q\leq 1$, $q>1$, or $q\leq
(1-4\,\pi^2)^{-1}$.

In the hyperbolic geometries, $S(\Theta(\eta))>1$ if
\be\l{eq93}
\cosh\Phi\,<\,\frac{1}{2\,\pi}\,\sqrt{\left |\frac{1-q}{q}\right|}\;,
\ee
i. e., near the singularity when $q>0$, or as $a\rightarrow\infty$ for $q<0$.
This can happen only for $0<q<(1+4\,\pi^2)^{-1}$  or $(1-4\,\pi^2)^{-1}<q<0$
(as $\cosh\Phi\geq 1$).

In order to solve (\ref{eq86}) we first make the substitution
\be\l{eq94}
Z=(\sin\Theta)^2
\ee       
and write
\be\l{eq95}
\mu(k,\eta)=\left(\frac{Z}{\vae}\right)^{q/2}\,f(Z)\;.
\ee
It is easily seen that the function $\mu$ defined by (\ref{eq95}) is a solution
of the wave equation (\ref{eq86}) as long as $f$ satisfies
\be\l{eq96}
Z\,(1-Z)\,\frac{d^2f}{dZ^2}+\left[\left(q+\frac{1}{2}\right)-(q+1)\,Z\right]\,
\frac{df}{dZ}+\frac{q^2}{4\,\vae}\,(k^2-\vae)\,f=0\;,
\ee
which is a hypergeometric equation \cite{abr} with parameters
\bea
\l{eq97}
{\cal A}&=&\frac{q}{2}\,\left(1+\frac{k}{\sqrt{\vae}}\right)\;,\\
\l{eq98}
{\cal B}&=&\frac{q}{2}\,\left(1-\frac{k}{\sqrt{\vae}}\right)\;,\\
\l{eq99}
{\cal C}&=&q+\frac{1}{2}\;.
\eea

The general solution of the gravitational wave equation is then given by
(\ref{eq95}) with $f(Z)$ being the general solution of the 
hypergeometric equation (\ref{eq96}). The explicit form of this solution depends
on several relations involving the parameters $\cal A$, $\cal B$, $\cal C$, and
may be found in \cite{abr}--\cite{erd}.
\section{Adiabatic vacuum solutions}
\l{s5}
\def\theequation{\thesection.\arabic{equation}}
\setcounter{equation}{0}
The next step is to find the particular solution which represents a `physical'
vacuum state. As it is usually done in the literature, we adopt the adiabatic
approach for defining particle states.
In the high-frequency limit the adiabatic definition reduces to the usual 
positive-frequency Minkowski modes ($\sim e^{-ik\eta}$).
A detailed discussion of this prescription can be found in \cite{birFul} and,
in our model, it  can be easily accomplished whenever
\be\l{eq100}
{\cal C}\neq 0,\,\pm 1,\,\pm 2,\,\ldots\;,
\ee
which is equivalent to
\be\l{eq101}
q\neq n-\frac{1}{2}\hspace{11mm},\hspace{3mm}n=0,\,\pm 1,\,\pm 2,\,\ldots\;,
\ee
or still
\be\l{eq102}
\gamma\neq\frac{2}{3}\,\frac{(2n+1)}{(2n-1)}\hspace{4mm},
\hspace{3mm}n=0,\,\pm 1,\,\pm 2,\,\ldots\;.
\ee
Under this restriction (which covers interesting cases such as $\gamma=0$,
$\gamma=4/3$, $\gamma=1$), the general solution of Eq. (\ref{eq96}) can be
written as \cite{abr}--\cite{erd}
\be\l{eq103}
f(Z)=C_1\,F({\cal A},\,{\cal B},\,{\cal C};\,Z)+C_2\,Z^{1-{\cal C}}\,
F({\cal A}^{'},\,{\cal B}^{'},\, {\cal C}^{'};\,Z)\;,
\ee
where $F$ stands for hypergeometric functions, $C_1$ and $C_2$ are arbitrary
constants, and
\bea
\l{eq104}
{\cal A}^{'}&=&{\cal A}-{\cal C}+1\;,\\
\l{eq105}
{\cal B}^{'}&=&{\cal B}-{\cal C}+1\;,\\
\l{eq106}
{\cal C}^{'}&=&2-{\cal C}\;.
\eea
The solution for the wave equation can then be expressed as
\be\l{eq107}
\mu(k,\eta)=C\,\left(\frac{Z}{\vae}\right)^{q/2}\,
\left[F({\cal A},\,{\cal B},\,{\cal C};\,Z) + C_0\,Z^{1-{\cal C}}\,
F({\cal A}^{'},\,{\cal B}^{'},\, {\cal C}^{'};\,Z)\right]\;.
\ee
The constant $C_0$ must be found by imposing the adiabatic constraints, whereas
$C$ is determined up to an irrelevant phase by the normalization condition
(\ref{eq60}). 

The fact that for both values of $\vae$, 
\bea
\l{eq108}
{\cal C}&=&{\cal A}+{\cal B}+\frac{1}{2}\;,\\
\l{eq109}
{\cal C}^{'}&=&{\cal A}^{'}+{\cal B}^{'}+\frac{1}{2}\;,
\eea
enables us to write the hypergeometric functions appearing in (\ref{eq107})
in terms of associated Legendre functions \cite{abr}--\cite{erd}. 

We anticipate the final result derived below and write the adiabatic solution
for the GW equation in the unified form
\be\l{eq110}
\mu(k,\eta) = e^{i\psi}\,e^{-i\,\frac{\pi}{4}\,\left(\frac{1-\vae}{2}\right)}
\;\Gamma(m+1)\left[\frac{|q|\,\Gamma(-m)}{4\,I\,\Gamma(m+1)}\right]^{1/2}\,
(\sin\Theta)^{1/2}\,{\cal L}(X)\;,
\ee
where:
\bea
\l{eq110b}
{\cal L}(X)&\equiv &{\cal P}^{-m}_{\nu}(X)\,-\,e^{\mp\, i\,m\,\pi\,\left(
\frac{1+\vae}{2}\right)}\:\frac{\Gamma(\nu-m+1)}{\Gamma(\nu+m+1)}\:
{\cal P}^m_{\nu}(X)\;,\\
\l{eq111}
X&\equiv &|\cos\Theta |\;,\\
\l{eq112}
m&\equiv &q-\frac{1}{2}=\frac{3}{2}\,\left(\frac{2-\gamma}{3\gamma-2}\right)\;,
\\
\l{eq113}
\nu&\equiv &\sqrt{\vae}\,|q|\,k-\frac{1}{2}\;,\\
\l{eq114}
I&\equiv &{\rm Im}\left\{e^{-\,i\,m\,\pi\,\left(\frac{1+\vae}{2}\right)}\:
\frac{\Gamma(\nu-m+1)}{\Gamma(\nu+m+1)}\right\}\;,
\eea
${\rm Im}\{z\}$ stands for the imaginary part of $z$, $\psi$ is a constant phase
(whose value is irrelevant for the evaluation of the Bogoliubov coefficients),
$\Gamma$ is the gamma function, and ${\cal P}^m_{\nu}$ represents the associated
Legendre function of the first kind, of order $m$ and degree $\nu$. The upper
(lower) sign applies if $\cos\Theta >0$ ($<0$). 

Note that for $\vae=-1$, $\cos\Theta=\cosh\Phi\geq 1$, but for $\vae=+1$, 
$\cos\Theta >0$ during an expansion phase if $q>0$ and during a contraction 
phase if $q<0$. We also have
\be\l{eq115}
0<Z\leq1\hspace{2mm},\hspace{2mm}0<X<1\hspace{12mm}\rm{(}\vae=+1{\rm )}
\ee
and
\be\l{eq116}
-\infty <Z=-(\sinh\Phi)^2<0\hspace{2mm},\hspace{2mm}1<X=\cosh\Phi<\infty
\hspace{7mm}{\rm (}\vae=-1{\rm )}\;.
\ee

In ref. \cite{grad} the `slanted' symbols ${\sl P}^m_{\nu}(X)$, 
${\sl Q}^m_{\nu}(X)$ are used to indicate the associated Legendre functions
of the first and second kind defined  for ${\rm Re}\{X\}>1$. The `straight'
symbols ${\sf P}^m_{\nu}(X)$, ${\sf Q}^m_{\nu}(X)$ are used for $|X|<1$. We
adhere to this notation whenever we are dealing explicitly with the cases
$\vae=-1$ and $\vae=+1$, respectively. The calligraphic symbols 
${\cal P}^m_{\nu}$ and ${\cal Q}^m_{\nu}$ will be used only in unified 
expressions like (\ref{eq110}) that are valid for both values of $\vae$. (Note
that, as it was remarked in \cite{allenCk95}, there is a misprint at page 999
of ref. \cite{grad}: the first paragraph should refer to `straight' $\sf P$
and the second paragraph to `slanted' $\sl P$.)

Some functional relations involving the associated Legendre functions (relations
with the hypergeometric functions, asymptotic behavior, derivatives, etc.),
which are needed to derive the adiabatic solution and the Bogoliubov 
coefficients, are distinct for ${\sf P}^m_{\nu}(X)$ and ${\sl P}^m_{\nu}(X)$
\cite{grad}. We must therefore treat the two cases separately.

For the elliptical models, and due to equations (\ref{eq108}) and (\ref{eq109}),
we can use Eq. (15.4.13) of ref. \cite{abr} to write
$F({\cal A},\,{\cal B},\,{\cal C};\,Z)$ and 
$F({\cal A}^{'},\,{\cal B}^{'},\, {\cal C}^{'};\,Z)$ in terms of ${\sf P}^{-m}
_{\nu}(X)$ and ${\sf P}^m_{\nu}(X)$, respectively. If we require $\mu(k,\eta)$
to have the desired asymptotic behavior for large $k$, the constant $C_0$ may
be determined. This can be done through the use of Eq. (8.10.7) of \cite{abr}.
In order to evaluate $C$ we apply the condition (\ref{eq60}) on the Wronskian
of the solutions by using Eq. (8.741.1) of \cite{grad}. Whenever we are forced
to deal with ${\sf P}^m_{\nu}(-\cos\Theta)$, we use Eq. (8.737.2) of 
\cite{grad}. The final result is then given by (\ref{eq110})--(\ref{eq114}) 
with $\vae=+1$ and ${\cal P}^m_{\nu}\mapsto {\sf P}^m_{\nu}$.

For the hyperbolical geometries, equations (\ref{eq108}) and (\ref{eq109})
enable us to use Eq. (15.4.12) of \cite{abr} to write
$F({\cal A},\,{\cal B},\,{\cal C};\,Z)$ and 
$F({\cal A}^{'},\,{\cal B}^{'},\, {\cal C}^{'};\,Z)$ in terms of ${\sl P}^{-m}
_{\nu}(\cosh\Phi)$ and ${\sl P}^m_{\nu}(\cosh\Phi)$, respectively. 
To analyze the
asymptotic behavior for large $k$ we make sucessive use of equations (8.723.1)
of \cite{grad} and (15.7.1) of \cite{abr}. The constant $C_0$ is then 
determined. For the Wronskian condition, we first write ${\sl P}^{-m}_{\nu}(X)$
as a combination of ${\sl P}^m_{\nu}(X)$ and ${\sl Q}^m_{\nu}(X)$ using
Eq. (8.736.1) of \cite{grad} and then apply (8.1.8) of \cite{abr}. We finally
get (\ref{eq110})--(\ref{eq114}) with $\vae=-1$ and ${\cal P}^m_{\nu}\mapsto
{\sl P}^m_{\nu}$.
\section{The Bogoliubov coefficients and the graviton spectrum}
\l{s6}
\def\theequation{\thesection.\arabic{equation}}
\setcounter{equation}{0}
\subsection{Review of the graviton creation mechanism}
The general procedure to calculate the graviton spectrum was explained in
detail in \cite{gw3t}. We shall give here only a brief resume of this method
and then evaluate the Bogoliubov coefficients for our present model. 

In realistic cosmological models the equation of state (\ref{eq15}) will change
its form during the cosmic evolution. Prior to a time $\er$ the equation reads
\be\l{eq117}
p_{(r-1)}(\eta)=(\gamma_{r-1}-1)\,\rho_{(r-1)}(\eta)\hspace{3mm},
\hspace{3mm}(\eta<\er)\;,
\ee
whereas for $\eta>\er$
\be\l{eq118}
p_{(r)}(\eta)=(\gamma_{r}-1)\,\rho_{(r)}(\eta)
\hspace{3mm},\hspace{3mm}(\eta>\er)\;.
\ee
We are working in the sudden transition approximation where the change in
$\gamma$ occurs instantaneously at the transition time $\er$. (See \cite{gw3t}
for the conditions of applicability of this approximation and other details.)
It is possible to impose the continuity conditions,
\be\l{eq119}
a_{(r-1)}(\er)=a_{(r)}(\er)\;,
\ee
\be\l{eq120}
a^{'}_{(r-1)}(\er)=a^{'}_{(r)}(\er)\;.
\ee
($g_{(r)}(\eta)$ indicates the appropriate form for the quantity $g$ in the
interval $[\er,\,\eta_{r+1}]$. The transition times are 
$\er$, $\eta_{r+1}$, etc. For the first phase we take $\eta_0$ to be any 
instant in that phase. The notation $g_r$ indicates $g_{(r)}(\er)$, etc.)

We write
\bea\l{eq122}
{\cal Y}_{(r-1)}(k,\eta)&=&\frac{\ms (k,\eta)}{a_{(r-1)}(\eta)}\hspace{10mm},
\hspace{3mm}\eta<\er\nonumber\\
 &=&\frac{1}{a_{(r)}(\eta)}\,\left[\ar\,\mr (k,\eta)+\br\,\mr^{*}(k,\eta)\right]
\hspace{3mm},\hspace{3mm}\eta>\er\;,
\eea
and 
\bea\l{eq123}
{\cal Y}_{(r)}(k,\eta)&=&\frac{1}{a_{(r-1)}(\eta)}\,\left[\alpha_r^{*}(k)\,
\ms(k,\eta)
-\br\,\ms^{*}(k,\eta)\right]\hspace{3mm},\hspace{3mm}\eta<\er\nonumber\\
 &=&\frac{\mr(k,\eta)}{a_{(r)}(\eta)}\hspace{10mm},\hspace{3mm}\eta>\er\;,
\eea
where $\ms$ and $\mr$ are solutions to the wave equation describing an adiabatic
vacuum in their respective regions. The two modes ${\cal Y}_{(r-1)}$ and 
${\cal Y}_{(r)}$ define two quantizations of the field associated with two
different Fock spaces, but the associated operators in each case represent
physical particle observables only inside their respective eras. It is easily
seen that
\be\l{eq124}
{\cal Y}_{(r-1)}(k,\eta)=\ar\,{\cal Y}_{(r)}(k,\eta)+\br\,
{\cal Y}_{(r)}^{*}(k,\eta)\;.
\ee
Due to (\ref{eq60}), the Bogoliubov coefficients \cite{birFul}, $\ar$, $\br$
satisfy
\be\l{eq125}
|\ar|^2-|\br|^2=1\;,
\ee
and the operators in the two epochs are related by a Bogoliubov transformation
\bea
\l{eq126}
A_{(r)}(\kk)&=&\ar\,A_{(r-1)}(\kk)+\beta_r^{*}(k)\,A^{\dagger}_{(r-1)}(-\kk)\;,
\\
\l{eq127}
A_{(r-1)}(\kk)&=&\alpha_r^{*}(k)\,A_{(r)}(\kk)-\beta^{*}_r(k)\,
A^{\dagger}_{(r)}(-\kk)\;.
\eea
Consequently, the vacuum state at the region $(r-1)$, labelled $|0_{(r-1)}>$,
is annihilated by $A_{(r-1)}$ but not by $A_{(r)}$, and, if $N_{(r)}(\kk)$ is
the number operator for the mode $\kk$ at stage $(r)$,
\be\l{eq128}
N_{(r)}(\kk)\equiv A^{\dagger}_{(r)}(\kk)\,A_{(r)}(\kk)\;,
\ee
then \cite{gw3t,birFul}
\be\l{eq129}
<N(\kk)>_r\equiv<0_{(r-1)}|N_{(r)}(\kk)|0_{(r-1)}>=|\br|^2\;.
\ee
Hence the vacuum state $|0_{(r-1)}>$ is a multi-particle state when we use the
definition of particles appropriate for the epoch $\eta>\er$. This is 
interpreted by saying that particles have been created by the expansion 
dynamics.

In a multi-stage model we need to relate the operators $A$ and $A^{\dagger}$
of stages separated by several transitions. This can be done recursively by
\cite{gw3t,gw1,allen}
\bea
\l{eq130}
\alpha_{T_r}(k)&=&\ar\,\alpha_{T_{r-1}}(k)+\beta_r^{*}(k)\,\beta_{T_{r-1}}(k)
\;,\\
\beta_{T_r}(k)&=&\br\,\alpha_{T_{r-1}}(k)+\alpha^{*}_r(k)\,\beta_{T_{r-1}}(k)
\;.
\eea

The Bogoliubov coefficients can be found if we impose the continuity of 
${\cal Y}_{(r-1)}$ and ${\cal Y}_{(r)}$ and their first derivatives at $\er$.
The result is \cite{gw3t,gw1}
\bea
\l{eq132}
\ar&=&i\,\left[\mr^{*}(k,\er)\,\ms^{'}(k,\er)-\ms(k,\er)\,\mr^{*'}(k,\er)\right]
\;,\\
\l{eq133}
\br&=&i\,\left[\ms(k,\er)\,\mr^{'}(k,\er)-\mr(k,\er)\,\ms^{'}(k,\er)\right]\;.
\eea

The power spectrum of the graviton background can be described by the quantity
$P_g(\omega)$, defined in such way that $P_g(\omega)\,d\omega$ represents the
energy per unit volume between the frequencies $\omega$ and $\omega+d\omega$
\cite{gw3t,gw1,allen}. In the units used in this paper 
\cite{gw3t,gw1,allen,starob}
\be\l{eq134}
P_g(\omega)=\frac{\omega^3}{\pi^2}\,<N_{\omega(k)}>\;.
\ee
The above expression is valid for waves with $\lambda\leq\lambda_H$. Modes
outside the Hubble radius ($\lambda>\lambda_H=H^{-1}$) cannot contribute to
the energy density since this is a locally-defined quantity (see 
\cite{gw3t,gw1,zelN83,allen,sahni90,zelN70} for details).

Note that equations (\ref{eq132})--(\ref{eq134}) are independent of the
cosmological model considered. Eq. (\ref{eq129}) assumes that no particles
were initially present. If the initial state is not the vacuum, then
`stimulated' graviton creation may also occur \cite{gw3t,birFul,ggv93}.
\subsection{The Bogoliubov coefficients}
In our model, the expressions for the Bogoliubov coefficients evaluated from
(\ref{eq110}), (\ref{eq132}), (\ref{eq133}) may be simplified if we use the
continuity conditions (\ref{eq119}), (\ref{eq120}), which read
\bea
\l{eq135}
\sin\Theta_{(r-1)}(\er)&=&\sin\Theta_{r-1}\,\left(\frac{a_r}{a_{r-1}}\right
)^{\frac{1}{q_{r-1}}}\;,\\
\l{eq136}
\cot\Theta_{(r-1)}(\er)&=&\amod\,\md\,\cot\Theta_r=\amod\,\frac{a_r\,H_r}
{\sqrt{\vae}}\;.
\eea
For $\vae=-1$, (\ref{eq135}) and (\ref{eq136}) reduce to
\bea
\l{eq137}
\sinh\Phi_{(r-1)}(\er)&=&\amd\,\sinh\Phi_{r-1}\,\left(\frac{a_r}{a_{r-1}}\right)
^{\frac{1}{q_{r-1}}}\;,\\
\l{eq138}
\coth\Phi_{(r-1)}(\er)&=&\amod\,\coth\Phi_r=\amod\,a_r\,H_r\;.
\eea

For $\vae=+1$ the final result is
\bea
\l{eq139}
\ar&=&i\,e^{i\,(\psi_{r-1}-\psi_r)}\:\left |\frac{q_{r-1}}{q_r}\right |^{1/2}\,
\frac{B_{r-1}\,B_r}{\pi}\,\sin\Theta_r\,\left(\frac{Y_r}{X_r}\right)^{1/2}\,
D_{\alpha_r}\;,\\
\l{eq140}
\br&=&i\,e^{i\,(\psi_{r-1}+\psi_r)}\:\left |\frac{q_{r-1}}{q_r}\right |^{1/2}\,
\frac{B_{r-1}\,B_r}{\pi}\,\sin\Theta_r\,\left(\frac{Y_r}{X_r}\right)^{1/2}\,
D_{\beta_r}\;,\\
\l{eq141}
X_r&=&|\cos\Theta_r|=\sin\Theta_r\,|a_r\,H_r|=\Omega_r^{-1/2}\;,\\
\l{eq142}
Y_r&\equiv &|\cos\Theta_{(r-1)}(\er)|=\frac{\sin\Theta_{r-1}}{\sin\Theta_r}\,
\left(\frac{a_r}{a_{r-1}}\right)^{\frac{1}{q_{r-1}}}\:X_r\;,\\
\l{eq143}
B_r&\equiv &\frac{|\Gamma(-m_r)|}{\Gamma(-m_r)}\,\frac{\Gamma(\nu_r-m_r+1)}
{\Gamma(\nu_r+m_r+1)}\,\left[-\frac{\sin(m_r\,\pi)}{I_r}\right]^{1/2}\;,\\
\l{eq144}
D_{\alpha_r}&\equiv &\pm D_r^{(1)}+i\,\frac{\pi}{2}\,D_r^{(2)}+\frac{\pi}{2}\,
\amd\,\md\,\left[\pm\frac{\pi}{2}\,D_r^{(3)}-i\,D_r^{(4)}\right]\;,\\
\l{eq145}
D_{\beta_r}&\equiv &\mp D_r^{(1)}+i\,\frac{\pi}{2}\,D_r^{(2)}+\frac{\pi}{2}\,
\amd\,\md\,\left[\pm\frac{\pi}{2}\,D_r^{(3)}+i\,D_r^{(4)}\right]\;,\\
\l{eq146}
D_r^{(1)}&\equiv &\frac{q_r}{q_{r-1}}\,\8\,\2-\4\,\6\;,\\
\l{eq147}
D_r^{(2)}&\equiv &\frac{q_r}{q_{r-1}}\,\8\,\1-\4\,\5\;,\\
\l{eq148}
D_r^{(3)}&\equiv &\frac{q_r}{q_{r-1}}\,\7\,\1-\3\,\5\;,\\
\l{eq149}
D_r^{(4)}&\equiv &\frac{q_r}{q_{r-1}}\,\7\,\2-\3\,\6\;,\\
\l{eq150}
l_r&\equiv &m_{r-1}\;,\\
\l{eq151}
n_r&\equiv &\nu_{r-1}\;.
\eea
The quantities $\Theta_r$, $\psi_r$, $q_r$, $m_r$, $\nu_r$, $I_r$ are those 
defined by equations (\ref{eq73}), (\ref{eq110})--(\ref{eq114}) appropriate to
the epoch $\er<\eta<\eta_{r+1}$ (and similarly for $\Theta_{r-1}$, etc.). The 
upper (lower) sign applies if $\cos\Theta_r >0$ ($<0$).

For $\vae=-1$ the coefficients can be written as
\bea
\l{eq152}
\ar&=&i e^{(\psi_{r-1}-\psi_r)} e^{-i\,(l_r+m_r)\,\pi}
\left(\frac{q_{r-1}}{q_r}\right)^{1/2}\frac{B_{r-1}\,B_r}{\pi}\sinh\Phi_r
\left(\frac{\overline{Y}_r}{\overline{X}_r}\right)^{1/2}
\overline{D}_{\alpha_r},\\
\l{eq153}
\br&=&ie^{(\psi_{r-1}+\psi_r)}e^{-i\,(l_r+m_r)\,\pi}
\left(\frac{q_{r-1}}{q_r}\right)^{1/2}\frac{B_{r-1}\,B_r}{\pi}\sinh\Phi_r
\left(\frac{\overline{Y}_r}{\overline{X}_r}\right)^{1/2}
\overline{D}_{\beta_r},\\
\l{eq154}
\overline{X}_r&\equiv&\cosh\Phi_r\;,\\
\l{eq155}
\overline{Y}_r&\equiv &\cosh\Phi_{(r-1)}(\er)=\frac{\sinh\Phi_{r-1}}
{\sinh\Phi_r}\,
\left(\frac{a_r}{a_{r-1}}\right)^{\frac{1}{q_{r-1}}}\,\overline{X}_r\;,\\
\l{eq156}
\overline{D}_{\alpha_r}&\equiv &\overline{D}^{(1)}_r+E_r\,
\overline{D}_r^{(2)}\;,\\
\l{eq157}
\overline{D}_{\beta_r}&\equiv &-\overline{D}_r^{(1)}\;,\\
\l{eq158}
E_r&\equiv &i\,\pi\,\frac{e^{i\,m_r\,\pi}}{\sin(m_r\,\pi)}\,\frac{{\rm Im}\{
\Gamma^{*}(\nu_r-m_r+1)\,\Gamma(\nu_r+m_r+1)\}}{\Gamma(\nu_r-m_r+1)\,
\Gamma^{*}(\nu_r+m_r+1)}\;,\\
\l{eq159}
\overline{D}_r^{(1)}&\equiv &\frac{q_r}{q_{r-1}}\,\I\,\B-\D\,\F\;,\\
\l{eq160}
\overline{D}_r^{(2)}&\equiv &\frac{q_r}{q_{r-1}}\,\I\,\A-\D\,\E\;.
\eea
%
\section{Concluding remarks}
\l{s7}
\def\theequation{\thesection.\arabic{equation}}
\setcounter{equation}{0}
We have addressed the problem of graviton creation in elliptical ($\vae=+1$) and
hyperbolic ($\vae=-1$) 
Friedmann-Robertson-Walker cosmologies with a gamma-law type
equation of state. After establishing the {\em general} solution for the scale
factor $a(\eta)$ we have obtained the {\em general} solution for the 
gravitational wave equation. We have then found those modes representing an
adiabatic vacuum state for $\gamma\neq\frac{2}{3}\,\left(\frac{2n+1}{2n-1}
\right)$, $n=0,\,\pm 1,\,\pm 2,\,\ldots$. This includes most cases of physical
interest, such as a de Sitter phase ($\gamma=0$), a radiation-dominated era
($\gamma=4/3$), and a dust phase ($\gamma=1$). By considering an arbitrary
sequence of epochs with $\gamma$ in this class, we have derived the 
Bogoliubov coefficients associated with graviton creation in this model. These
coefficients are necessary to obtain the gravitational wave power spectrum,
$P_g(\omega)$, and to give an accurate determination of the graviton
contribution to the anisotropy of the CMB \cite{allenK94}.

The application of our results to a definite model, that is, to a definite
sequence of cosmic eras corresponding to particular values of $\gamma$, can, in
principle, be used to compare the behavior of the relic gravitons in the three
possible geometries ($\vae=0,\;\pm 1$). In fact, by imposing the 
{\em observational} constraints associated with the anisotropy of the CMB, with
the nucleosynthesis of $^4{\rm He}$, and with the limits on the present value
of $\Omega$, it may be possible to obtain {\em theoretical} constraints 
involving
several cosmological parameters, such as the value of $\gamma$ in an initial
inflationary period, the value of $H$ at the end of inflation, and $\Omega$ 
itself. 

Some other interesting questions also arise. For example, in elliptical models,
what happens to graviton creation as the expansion maximum is reached? And, 
for models with a big-bounce 
($q<0$), what can be said about this production near the expansion minimum? 
\cite{jdb} Some of these issues will be addressed in a forthcoming communication
\cite{maiaL}.
\section*{Acknowledgments}

It is a pleasure to thank 
Robert Brandenberg 
and Adolfo Maia for valuable suggestions.
This work was partially supported by the Conselho Nacional de Desenvolvimento Cient\`{i}fico e Tecnol\'{o}gico-CNPQ (Brazilian Research Agency)
and by the US Department of Energy under grant 
DE-F602-91ER40688, Task A.


\end{document}